\documentclass[prd, twocolumn, nofootinbib, floatfix]{revtex4-1}

\usepackage{amsmath}
\usepackage{graphicx}
\usepackage{dcolumn}
\usepackage{bm}
\usepackage{epsfig}
\usepackage{amssymb,latexsym,mathrsfs}
\usepackage{graphicx}
\usepackage{color}
\usepackage{hyperref}

\hypersetup{
    colorlinks=true,
    linkcolor=red,
    citecolor=blue,
} 

\newcommand{\be}{\begin{equation}}
\newcommand{\ee}{\end{equation}}
\newcommand{\bs}{\begin{split}} 
\newcommand{\bea}{\begin{eqnarray}}
\newcommand{\eea}{\end{eqnarray}}

\newcommand{\lcdm}{$\Lambda$CDM} 
\newcommand{\gm}{G_{\rm matter}} 
 
\newcommand{\gl}{G_{\rm light}}

\newcommand{\al}{\alpha}

\newcommand{\clbb}{C_\ell^{BB}}
\newcommand{\clpp}{C_\ell^{\phi\phi}}
\newcommand{\hic}{{\tt hi\_class}} 
\newcommand{\alens}{A_{\rm lens}}

\begin{document}

\title{No Slip CMB} 

\author{Micah Brush${}^1$, Eric V.\ Linder${}^{1,2}$, Miguel Zumalac{\'a}rregui${}^{1,3}$} 
\affiliation{${}^1$Berkeley Center for Cosmological Physics \& Berkeley Lab, 
University of California, Berkeley, CA 94720, USA\\ 
${}^2$Energetic Cosmos Laboratory, Nazarbayev University, 
Astana, Kazakhstan 010000\\ 
${}^3$Institut de Physique Th\' eorique, Universit\'e  Paris Saclay 
CEA, CNRS, 91191 Gif-sur-Yvette, France}

%%%%%%%%%%%%%%%%%%%%%%%%%%%%%%%%%%%%%%%%%%%%%%%%%%%%%%%%%%%%%%%%%%%%%%%%
\begin{abstract} 
No Slip Gravity is a simple modified gravity theory with only one free 
function and the interesting characteristic that -- unlike many modified 
gravity theories -- it suppresses growth. This allows it to fit current 
redshift space distortion and $\sigma_8$ mass fluctuation amplitude data 
better than $\Lambda$CDM in general relativity, while retaining a \lcdm\ 
background expansion and hence distances. Since it has no gravitational slip it alters equally
CMB lensing and matter density growth, and in addition affects the CMB gravitational wave B-mode polarization power spectrum. 
We investigate and compute the effects of No Slip Gravity for CMB lensing 
and B-modes, and present a simple analytic approximation. Using a Monte Carlo 
analysis, we place constraints on the theory from current CMB data. 
\end{abstract}

\date{\today} 

\maketitle

%%%%%%%%%%%%%%%%%%%%%%%%%%%%%%%
\section{Introduction}

Acceleration of the cosmic expansion may be caused by a cosmological 
constant or scalar field, or a modification of the laws of gravity. 
Once the expansion history is given, general relativity implies that the 
growth of cosmic structure is determined, but modified gravity allows 
deviations to exist and so offers further signatures. If cosmic growth is indeed lower than expected from 
general relativity plus a background expansion near a cosmological constant 
and cold dark matter (\lcdm) universe, as several current data sets 
from redshift space distortions and weak gravitational lensing may imply 
\cite{des,des2,kids,leauthaud,boss}, 
then modified gravity is an interesting possibility to examine. 

Generally scalar-tensor theories of gravity lead to an enhancement of 
growth, however, due to the scalar field adding to the gravitational 
strength. A theory with a strong braiding between the scalar and tensor 
sectors can counteract this, and even weaken gravity and growth. 
One interesting example of such a theory is No Slip Gravity \cite{nsg}. 
This also has the useful property that there is only one (out of a generic 
four) modified gravity function (formally, property functions \cite{bellini}) 
that must be specified. Thus it can be a fairly predictive theory. 

Since this one free function, which can be regarded as the running of 
the Planck mass, also affects the propagation of gravitational waves 
(even though their speed is fixed to the speed of light), No Slip Gravity 
is of particular interest for experiments seeking primordial gravitational 
waves signatures from inflation in the cosmic microwave background (CMB) 
B-mode polarization power spectrum. In addition, \cite{nsg} showed that 
the modified propagation of gravitational waves from late universe sources 
such as could be detected by ground and space based gravitational wave 
interferometers must be directly related to an offset in the distances 
to a source derived from gravitational wave emission vs from photons -- and to structure growth -- 
a testable prediction. 

Finally, the suppression of cosmic structure growth 
that would be seen in galaxy clustering surveys should also be directly 
related to a dilution in the CMB lensing power spectrum, due to the no 
slip nature of the theory, i.e.\ that the gravitational strength for 
influencing matter and light are equal. This implies the small angular 
scale part of the CMB B-mode polarization power spectrum, dominated by 
lensing, should be affected. 

This tight interrelation between many observables, whose measurements 
will be increasingly precise with the next generation of galaxy clustering, 
CMB, weak lensing, and gravitational wave experiments, makes No Slip 
Gravity well worth testing. Here we used \hic, a Boltzmann code enhanced 
to include modified gravity property functions in the equations, to compute 
the CMB power spectra, especially lensing and B-modes, and investigate the 
signatures of No Slip Gravity. 

In Section~\ref{sec:method} we briefly review modified gravity property 
functions, No Slip Gravity, and the \hic\ code. We evaluate the CMB B-mode 
polarization power spectrum in Sec.~\ref{sec:bmode} and elucidate the 
physical effects on the tensor (gravitational wave) part and the lensing 
part. Section~\ref{sec:lensing} investigates in greater detail the CMB 
lensing power spectrum and gives a simple analytic approximation for the 
effects. We carry out a Markov Chain Monte Carlo 
analysis of the constraints from current CMB data in Sec.~\ref{sec:mcmc}, and discuss the results 
and conclude in Sec.~\ref{sec:concl}.

%%%%%%%%%%%%%%%%%%%%%%%%%%% 
\section{Methods for Theory and Code} \label{sec:method} 

The property functions of modified gravity \cite{bellini} are an effective 
field theory like approach to characterizing deviations from general relativity, 
with each property function closely related to a particularly characteristic. 
In standard Horndeski theory, describing scalar-tensor gravity that yields 
second order equations of motion, four property functions enter: $\al_K$ 
giving the kinetic stiffness of the scalar field fluctuations, $\al_B$ the kinetic mixing (or braiding) 
between the scalar field and the metric, $\al_M$ the running of the effective Planck mass (on the cosmological background), and $\al_T$ the deviation of the speed 
of tensor perturbation (gravitational wave) propagation from the speed of 
light. 
Each is a dimensionless 
function of time, or cosmic scale factor, and their vanishing indicates that 
gravity is general relativity. 

Certain subclasses of Horndeski gravity have relations between the property 
functions, with the two main examples being $f(R)$ gravity, Brans-Dicke, or chameleon theories where $\al_B=-\al_M$ and No Slip Gravity where $\al_B=-2\al_M$. It is the condition  
\be 
\al_B=-2\al_M \quad({\rm No\ Slip\ Gravity}) 
\ee 
that gives the physical characteristic of no slip, i.e.\ the metric potentials 
are equivalent (neglecting matter anisotropic stress) and so the 
gravitational coupling strengths $\gm$ and $\gl$ (generalizations of Newton's constant) 
that enter the modified Poisson equations for density perturbations and 
light propagation are equal. 

As mentioned in the Introduction, No Slip Gravity has some interesting 
physical properties in addition to its mathematical simplicity, with an 
intriguing suppression of growth. It can also readily be made ghost free, 
satisfying $\al_K+(3/2)\al_B^2\ge0$ (see the discussion of $\al_K$ below), 
and has a remarkably simple stability condition, 
\be 
\frac{d^2\ln M_\star^2}{dt^2}\le0 \quad{\rm or}\quad 
(H\al_M)\,\dot{}\le0 \ , 
\ee 
where $M_\star$ is the effective Planck mass, $H$ the 
Hubble parameter, and dot denoting a time derivative. Indeed, for certain 
parametrizations of $\al_M(a)$, No Slip Gravity is an exemplar, a 
``maximally stable'' model \cite{stable}. 

We follow the parametrizations for $\al_M(a)$ given in \cite{nsg}, that 
were stable and ghost free. While we tested both the $M_\star$ and $\al_M$ 
parametrizations given there, we found them substantially equivalent and 
so we present the results for 
\bea
\label{eq:am}
\al_M(a)&=&c_M\left(1-\tanh^2[(\tau/2)\ln(a/a_t)]\right)\\ 
&=&\frac{4c_M\,(a/a_t)^\tau}{[(a/a_t)^\tau+1]^2} \ . 
\eea  
Note that $\al_M$ approaches zero in the past (general relativity at high 
redshift) and in the future (consistent with the de Sitter asymptote of 
\lcdm), and has a maximum value $c_M$ at $a=a_t$. The stability condition is just 
$\tau\le 3/2$. We set fiducial values $a_t = 0.5$, $\tau = 1$, and $M_{\star,{\rm ini}}^2=m_p^2$, which are used for the rest of the paper.

We set $\al_T=0$, since this is the most straightforward interpretation of the 
GW170817/GRB170817A gravitational wave and electromagnetic counterpart observations \cite{ligo}. 
The kineticity $\al_K$ generally has little effect on observational quantities 
on subhorizon scales (see the discussion below). 
We take the background expansion history to be that of $\Lambda$CDM so 
that we can compare the effects on the CMB and growth purely due to the 
gravitational deviations from general relativity.

The model was introduced into the \hic\ code \cite{Zumalacarregui:2016pph,Blas:2011rf} by imposing the No Slip Gravity condition $\al_B = -2 \al_M$, with the specific time dependence given by Eq.~(\ref{eq:am}). 
The effective Planck mass $M_\star$ is obtained from the definition $\al_M \equiv d\ln M_\star^2/d\ln a$.
These model-dependent specifications are included in the \texttt{background} module of \hic\ and are read automatically by the \texttt{perturbations} module to solve the linearly-perturbed, modified Einstein equations for Horndeski gravity. A public version of \hic\ including these modifications is available at \url{https://github.com/micbru/hi_class_public}. It can also be used to specify $\alens$ (as discussed in Secs.~\ref{sec:lensing}and~\ref{sec:mcmc}).

To test the influence of $\al_K$, and in particular to separate out its  
effect in No Slip Gravity, we studied the results for several different 
values approaching $\al_K = 0$. 
One cannot generally set $\al_K = 0$ identically because the inverse of the kinetic term $\al = \al_K + (3/2)\al_B^2$ diverges in the limit in which $\al_B\to 0$, as happens at early times. 
This divergence produces very rapid oscillations of the scalar field perturbations that cannot be resolved numerically. Furthermore, these oscillations have no impact on the predictions in the limit in which $\Omega_\phi, \al_B, \al_M$ are all negligible, which is satisfied at early times in the models we are considering.

\begin{figure}[htbp!]
\includegraphics[width=\columnwidth]{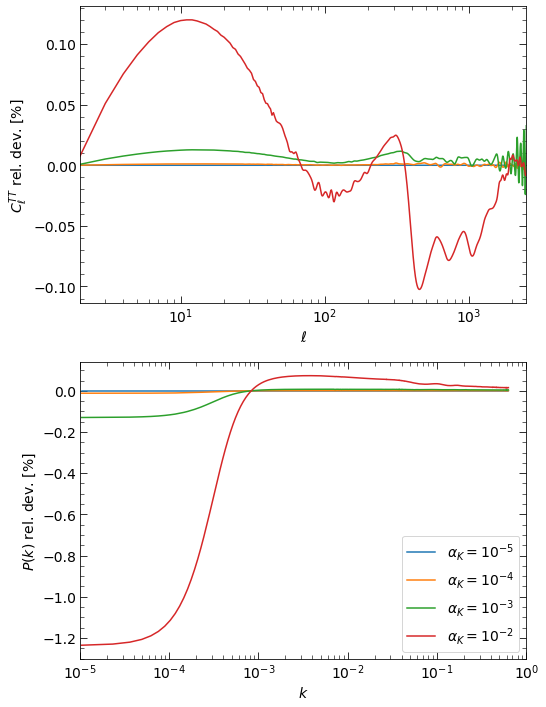}
\caption{
The CMB temperature power spectrum $C_\ell^{TT}$ and the matter density perturbation power spectrum $P(k)$ compared with different values of constant $\al_K$ relative to $\al_K=10^{-5}$. The deviations are 
well under a percent, with only the matter power spectrum showing larger 
deviations for $\al_K\gtrsim 10^{-2}$ on near horizon scales. $c_M$ is held fixed at 0.1. 
} 
\label{fig:smallak} 
\end{figure}

To overcome these numerical problems at early times we add a constant value to $\al_K$ through the \texttt{kineticity\_safe\_smg} parameter in \hic. 
We tested that this value is small enough to produce a negligible deviation in the CMB spectra:
Figure~\ref{fig:smallak} compares the CMB temperature power spectrum 
$C_\ell^{TT}$ and the matter density perturbation power spectrum $P(k)$ 
for different values of this parameter while holding the other parameters fixed. 
This value is several orders of magnitude below the sensitivity of even next-generation experiments \cite{Alonso:2016suf}.

%%%%%%%%%%%%%%%% 

Indeed, $\al_K$ has little influence on subhorizon scales (wavenumbers 
$k\gtrsim 10^{-3}\,h$/Mpc). For scales near the horizon,  
as long as $\al_K\lesssim  10^{-3}$ the effect on the power spectra is negligible. We set it to $10^{-5}$ for the rest of the computations.

To ensure that small \texttt{kineticity\_safe\_smg} works as expected, we also made $\al_K$ near zero by setting $\al_K = 10^{-4} \alpha_M$. There was no discernible difference between these two methods for setting $\al_K = 0$ and we believe them to be essentially equivalent at the level of precision of both modified gravity Boltzmann codes and current data \cite{Bellini:2017avd}.

%%%%%%%%%%%%%%%%%%%%%%%%%% 
\section{CMB B-Mode Polarization} \label{sec:bmode} 

An observable of great interest for future surveys is the B mode 
polarization power spectrum of the CMB. This has two main contributions: 
at large angular scales, low multipoles $\ell$, the B modes arise 
from the primordial gravitational waves generated by inflation in the 
early universe. Their amplitude is a direct measure of the inflationary energy scale. 
%\mz{Reference?} 
Furthermore, they test the theory of gravity 
because they involve the propagation of gravitational waves over 
cosmic distances. At small angular scales, high multipoles, the dominant 
B mode contribution comes from gravitational lensing of the E mode 
polarization and so does not substantially involve the tensor 
perturbations. However, the strength of gravitational lensing depends 
on the matter density perturbation power spectrum, and hence the growth of cosmic 
structure, and also the gravitational coupling strength for light 
propagation. Thus it tests both structure growth and gravity. 

\begin{figure}[htbp!]
\includegraphics[width=\columnwidth]{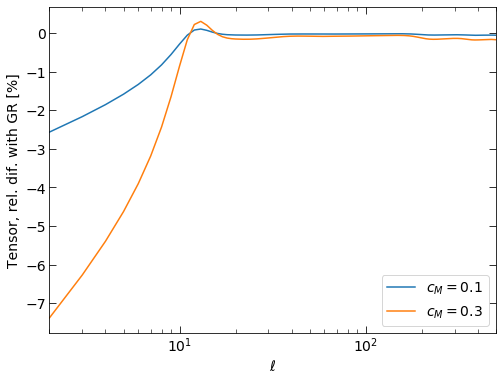} 
\caption{
The primordial CMB B modes give signatures of modified gravity, even if the 
gravitational wave speed is the speed of light. The relative deviation 
of the tensor power spectra $C_\ell^{BB}$ (without lensing) for No Slip Gravity models with $c_M=0.1$ and 0.3 is shown 
with respect to general relativity ($c_M=0$).} 
\label{fig:bonly} 
\end{figure}

For primordial gravitational waves, the main signatures are a 
reionization bump at $\ell\lesssim 10$ and a recombination bump at 
$\ell\approx 100$ in the B mode spectrum. Since the gravitational wave 
propagation speed is unchanged from general relativity (since $\al_T=0$) 
then the angular scale of the bumps are unchanged by the modified 
gravity, just the amplitude. 
For primordial tensor to scalar ratio $r\lesssim0.01$, the recombination 
bump is mostly hidden beneath the late time lensing contribution. 
Figure~\ref{fig:bonly} shows the tensor only (no lensing) contributions to the B modes with $r=0.1$
for different amplitudes $c_M$, relative to the general relativity 
case.

%%%%%%%%%%%%%%%% 

The manner in which $\al_M$ enters into gravitational wave propagation 
is through the friction term in the wave propagation (see e.g.\ \cite{amendola,1408.2224}). Positive $\al_M$ damps the gravitational wave amplitude. This can also be thought of as increasing the luminosity distance of the gravitational wave source \cite{nishizawa,arainish}. 
Thus the B mode power is reduced. The rule of thumb for this No Slip 
Gravity model is that the reionization bump is reduced by 
\be 
\Delta P_{BB}/P_{BB}\approx 20\,c_M\,\% \ . 
\ee 
In contrast, the recombination bump is quite insensitive to $c_M$ in this model. 
This is because $\al_M$ is time dependent and is much smaller
at $z\approx 1000$ than at $z\approx 6$.
Note that \cite{stable} used a model where 
$\al_M=c_M\,a$ and in their Figure~4 found a comparable effect on the 
B mode reionization bump, when accounting for the time dependence and 
different value of $c_M$. 

Since the reionization bump  amplitude is 
proportional to the tensor-scalar ratio $r$, there is a degeneracy between the inflationary and modified gravity signatures if we only detect the reionization bump. The bias is
\be 
\delta r/r \approx 0.2\,c_M \ . 
\ee 
Since we expect $c_M\approx 0.03$ (at least if it is connected 
with the suppression of growth at the current data level), this is less than a 1\% bias. If $r=0.01$, then even for Stage 4 
CMB experiments the bias will not be significant. Moreover, the higher multipole B mode spectrum will 
have a separate modified gravity signature that could break this degeneracy (though it may involve other covariances, as we will see in Sec.~\ref{sec:mcmc}).

\begin{figure}[tbp!]
\includegraphics[width=\columnwidth]{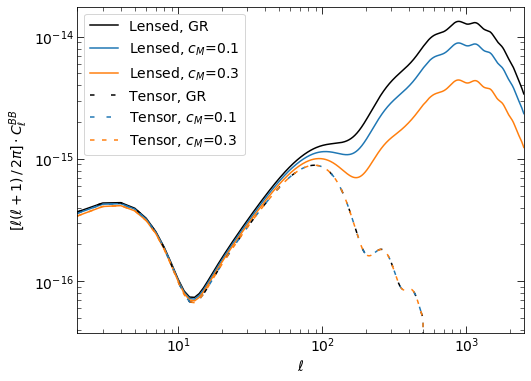} 
\caption{
CMB B mode power spectra including both lensing and primordial tensor (with $r=0.1$) contributions are plotted for the No Slip Gravity models 
with $c_M=0.1$ and 0.3, and general relativity ($c_M=0$). The dotted 
curves show the same models but for the tensor contributions alone. 
} 
\label{fig:bmodes} 
\end{figure}

We now turn to the overall B mode spectrum, including both the tensor 
contributions (with $r=0.1$ to make the 
recombination bump clearly visible) and lensing contributions. 
Figure~\ref{fig:bmodes} shows the full B mode spectrum where we can see the 
two primordial bumps plus the lensing peak at $\ell\approx 1000$. The 
effect of $\al_M$ on the lensing power is much more pronounced there.

%%%%%%%%%%%%%%%% 

\begin{figure}[tbp!]
\includegraphics[width=\columnwidth]{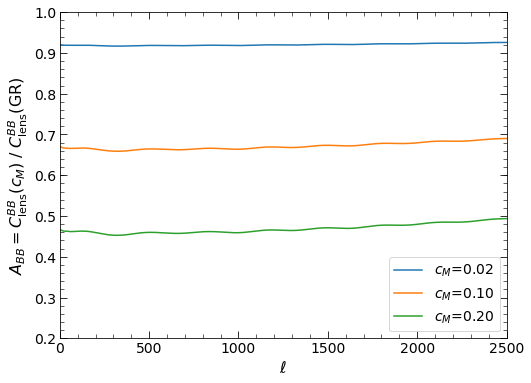} 
\caption{
Ratios of the lensed B modes (no tensor contribution) in No Slip Gravity for the indicated values of 
$c_M$ relative to GR.  
} 
\label{fig:bbr0} 
\end{figure}

Since a significant effect is visible in the high multipoles from the 
CMB lensing, we now focus on the CMB B mode lensing by setting $r=0$. 
Figure~\ref{fig:bbr0} shows the ratio of the scalar lensed $\clbb$ 
between No Slip Gravity and general relativity for various $c_M$. Increasing 
$\al_M$ suppresses the power of the lensing B modes as it weakens the 
growth and matter power spectrum.

%%%%%%%%%%%%%%%% 

The lensing B modes exhibit a substantially scale independent dilution, 
given approximately by 
\be 
A_{BB}\equiv \frac{C^{BB}_{\rm lens}(c_M)}{C^{BB}_{\rm lens}(GR)}\approx 1-4c_M \ , 
\ee 
for $c_M\ll1$. We explore the magnitude of $A_{BB}$ further in the next section. 

\begin{figure}[tbp!]
\includegraphics[width=\columnwidth]{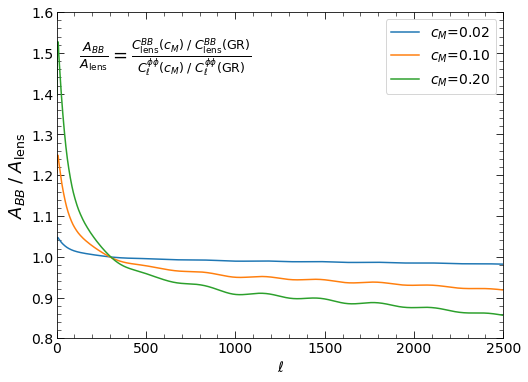} 
\caption{
Ratio of ratios: the lensing B mode power spectrum relative to the CMB lensing power 
spectrum is compared between the No Slip Gravity cases and the general 
relativity prediction, or equivalently the ratio of lensing B modes in 
No Slip Gravity relative to general relativity ($A_{BB}$) is compared 
to the ratio of lensing power in No Slip Gravity relative to general 
relativity ($\alens$).
} 
\label{fig:bbppr0} 
\end{figure}

Since the lensing B modes arise from gravitational lensing (affected 
by the late time modified gravity) of the primordial E modes (unaffected by the 
late time modified gravity), it is useful to directly compare the CMB lensing 
deflection power spectrum. The ratio of 
$C^{\phi\phi}_L(c_M)/C^{\phi\phi}_L(GR)$ is called $A_{\rm lens}$ and is not 
identical to $A_{BB}$ due to a convolution going from $C^{\phi\phi}$ 
to $C^{BB}_{\rm lens}$. Figure~\ref{fig:bbppr0} illustrates how the ratio behaves.

%%%%%%%%%%%%%%%% 

In this case scale dependence is evident. There is an enhancement in 
the ratio $A_{BB}/A_{\rm lens}$ at low $\ell$ and a dilution at high $\ell$. 
We explore the origin of this in the next section.

%%%%%%%%%%%%%%%%%%%%%%%%% 
\section{CMB Lensing Power Spectrum} \label{sec:lensing} 

To separate the effects seen in the previous section about the effects 
of No Slip Gravity on the B mode power spectrum (dominated by lensing) 
and the CMB lensing power spectrum itself, we now investigate the CMB lensing power spectrum. This comes from the power spectrum of the gravitational potential along the line of sight, $\clpp$. 

\begin{figure}[tbp!]
\includegraphics[width=\columnwidth]{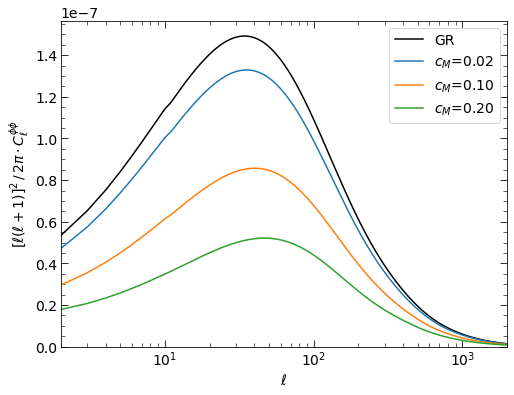} 
\caption{
The CMB lensing power spectra (or deflection angle power with the 
extra factors of $\ell$) are plotted for No Slip Gravity for various 
$c_M$ and for general relativity ($c_M=0$).} 
\label{fig:phiphi} 
\end{figure}

Figure~\ref{fig:phiphi} plots $\clpp$ in No Slip Gravity for various 
$c_M$ as well as in general relativity. As expected, the gravitational 
potential power is diminished as structure growth is suppressed by 
increasing $c_M$. We have also studied the effect of varying $a_t$ 
and $\tau$ from their fiducial values, 
but found that the effects were fairly degenerate with changing $c_M$. 
The key quantity seems to be the integrated amplitude of the modified gravity 
deviation over the redshift range during which CMB lensing has the 
greatest effect, $z\approx 1-3$. We give an analytic justification for 
this numerical finding below.

%%%%%%%%%%%%%%%% 

\begin{figure}[tbp!]
\includegraphics[width=\columnwidth]{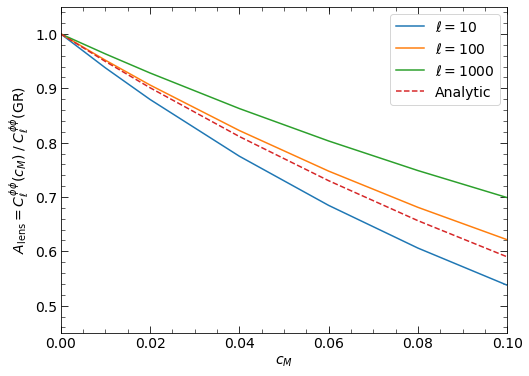} 
\caption{
The ratio of the CMB lensing power spectrum between No Slip Gravity 
and general relativity, as a function of modified gravity strength 
$c_M$, gives a clear idea of the physical growth suppression. Different 
solid color curves represent different scales $\ell$. The dashed 
curve is the approximate analytic result from Eq.~\ref{clppapx}.
}

\label{fig:phiphimu} 
\end{figure}

To show the impact of No Slip Gravity more directly, and the scaling 
with $c_M$ as a function of scale, we plot the deviation of the lensing 
power $\clpp$ from the general relativity result as a function of 
$c_M$ for several multipoles $\ell$ in Fig.~\ref{fig:phiphimu}. 
$\clpp$ can exhibit significant deviations from general relativity, 
depending on $c_M$ and to a lesser extent $\ell$, i.e.\ there is some scale dependence.

%%%%%%%%%%%%%%%% 

We can explore the effect on the CMB lensing power spectrum in a 
heuristic, analytic manner. A key characteristic of No Slip Gravity is 
that matter and light feel the same gravitational strength. In terms of 
the modified Poisson equation gravitational strengths (dimensionless, i.e.\ in units of 
Newton's constant $G_N$), 
\be 
\gl=\gm=\frac{m_p^2}{M^2_\star} \ , \label{eq:nsggm}
\ee 
where the first equality is an expression of no slip, and the second 
follows from the $\al_B=-2\al_M$ relation in No Slip Gravity 
\cite{nsg}. 

In illustrative terms, the CMB lensing power spectrum 
\be 
\clpp\sim [\nabla^2(\Phi+\Psi)]^2\sim \gl^2 (\delta\rho/\rho)^2 \ , 
\ee 
where $\Phi$ and $\Psi$ are the time-time and space-space metric 
potentials, $\gl$ is the modified gravitational strength in the 
relativistic Poisson equation 
\be 
\nabla^2(\Phi+\Psi)=8\pi G_N\, \gl\, (\delta\rho/\rho) \ , 
\ee 
and $\delta\rho/\rho$ is the linear matter density contrast. 

If we now take the ratio of $\clpp$ to that from general relativity 
\lcdm, we have 
\be 
\frac{\clpp}{C_{\ell}^{\phi\phi} ({\rm GR})}\approx \gl^2 \left(1+2\frac{\delta g}{g_{\rm GR}}\right) \ , 
\ee 
where $g=(\delta\rho/\rho)/a$ is the density perturbation growth factor and $\delta g$ is the (small) 
difference from the general relativity case. Since much of the contribution to 
CMB lensing occurs at redshift $z\approx3$, we can use the modified gravity 
growth approximation of \cite{mgarea} that relates the early time modified growth 
to 3/5 times the area under the modified gravitational strength function, 
Area$\,\sim\int d\ln a\,(\gm-1)$. This is intended as a rough rule of thumb. 

Putting this together, we find 
\bea 
\frac{\clpp}{C_{\ell}^{\phi\phi} ({\rm GR})}&\approx& \left(\frac{m_p^2}{M^2_\star}\right)^2  
\left[1+\frac{6}{5}{\rm Area}\right]\\ 
&\approx& \left(\frac{m_p^2}{M^2_\star}\right)^2 
\left[1+\frac{6}{5}\int d\ln a\,\left(\frac{m_p^2}{M^2_\star}-1\right)\right] 
\eea 
Finally we use the No Slip Gravity model expression for $M^2_\star$ \cite{nsg},  
\be 
\frac{m_p^2}{M^2_\star}=e^{-(2c_M/\tau)[1+\tanh((\tau/2)\ln(a/a_t)]} \ , 
\ee 
and evaluate it near the peak $a\approx a_t$, using a ``full width half maximum'' weighting on the integral to find 
\be 
\frac{\clpp(c_M)}{C_{\ell}^{\phi\phi} ({\rm GR})}\approx e^{-4c_M/\tau}\,\left(1-\frac{6c_M}{5}\right) \ . \label{clppapx} 
\ee 
This is shown as the dashed ``Analytic'' curve in Fig.~\ref{fig:phiphimu}, and works surprisingly well at multipoles near the maximum of the lensing power spectrum. 
For small amplitudes $c_M\ll1$ this gives a lensing power spectra ratio 
of $\approx1-5c_M$. Note that while this ratio is basically $A_{\rm lens}$, 
in fact $A_{\rm lens}$ is usually taken to be scale independent. 

The lensing B mode power is a convolution of the lensing power 
spectrum with the E mode polarization power spectrum. The power 
$\clbb$ at some multipole actually arises from a range of multipoles 
in $C^{\phi\phi}_{\ell'}$. This tends to weaken the scale dependence 
of $\clbb/C^{BB}_{\rm lens} ({\rm GR})$ as seen in Fig.~\ref{fig:bbr0}.

We can 
also gain intuition on Fig.~\ref{fig:bbppr0}. At low multipoles, the 
convolution prefers higher lensing multipoles, shifted up toward the 
peak of the lensing power spectrum, while at high multipoles it 
prefers somewhat lower lensing multipoles, again toward the peak. But 
from Fig.~\ref{fig:phiphimu} we see that shifting low lensing multipoles 
to higher ones relaxes the dilution suffered by the lensing power, while 
shifting high lensing multipoles to lower ones increases the dilution. 
Thus, at low multipoles $A_{BB}>A_{\rm lens}$ and at high multipoles $A_{BB}<A_{\rm lens}$ as found in Fig.~\ref{fig:bbppr0}. Current uncertainties on $A_{BB}$ 
are large, with the POLARBEAR CMB experiment quoting 
$A_{BB}=0.69^{+0.26}_{-0.25}({\rm stat}^{+0}_{-0.04})({\rm inst}\pm0.04$) \cite{polarbear}. While 
POLARBEAR's estimate of $A_{\rm lens}$ is slightly lower, Planck's much more 
precise value is $A_{\rm lens}=1.180\pm0.065$ \cite{planck6}, i.e.\ $A_{BB}<A_{\rm lens}$. However it is much too early to attribute any significance to 
this.

%%%%%%%%%%%%%%%%%%%%%%%%% 
\section{Current Constraints} \label{sec:mcmc} 

To explore CMB constraints on No Slip Gravity, in the form of the effective $c_M$ amplitude of the modified gravity deviation (i.e.\ folding the $a_t$ and $\tau$ parameters into a single effective amplitude parameter) we run a Markov Chain Monte Carlo Analysis using the MontePython code \cite{Audren:2012wb,Brinckmann:2018cvx} interfaced with \hic. 
Precision B mode data does not yet exist so we use only the
\texttt{Plik} TT likelihoods from the Planck 2015 release \cite{planck11}. 
The parameter $c_M$ is added to the 
six standard cosmology parameters, with a stability restriction $c_M\ge0$. 

We find that the 95\% confidence upper limit on $c_M$ is 0.015. Such a tight 
limit arises from two effects. As we have seen, $c_M$ suppresses the lensing 
power spectrum, while Planck data prefer a higher than $\Lambda$CDM strength 
of lensing, i.e.\ smoothed acoustic oscillations corresponding to $\alens=1.18$ \cite{planck6}. Secondly, $c_M$ gives a larger 
Sachs-Wolfe effect on large angles (low multipoles), again in tension with what 
the data suggest. We can separate out these effects by looking at $\ell>30$ data 
vs $\ell<30$ data and find that each gives roughly the same $\Delta\chi^2\sim10$ penalty to a value of $c_M=0.03$ with respect 
to general relativity ($c_M=0$). 
Recall that conversely at least some large scale structure data 
prefer suppressed growth as discussed in \cite{nsg}. 

\begin{figure}[tbp!]
\includegraphics[width=\columnwidth]{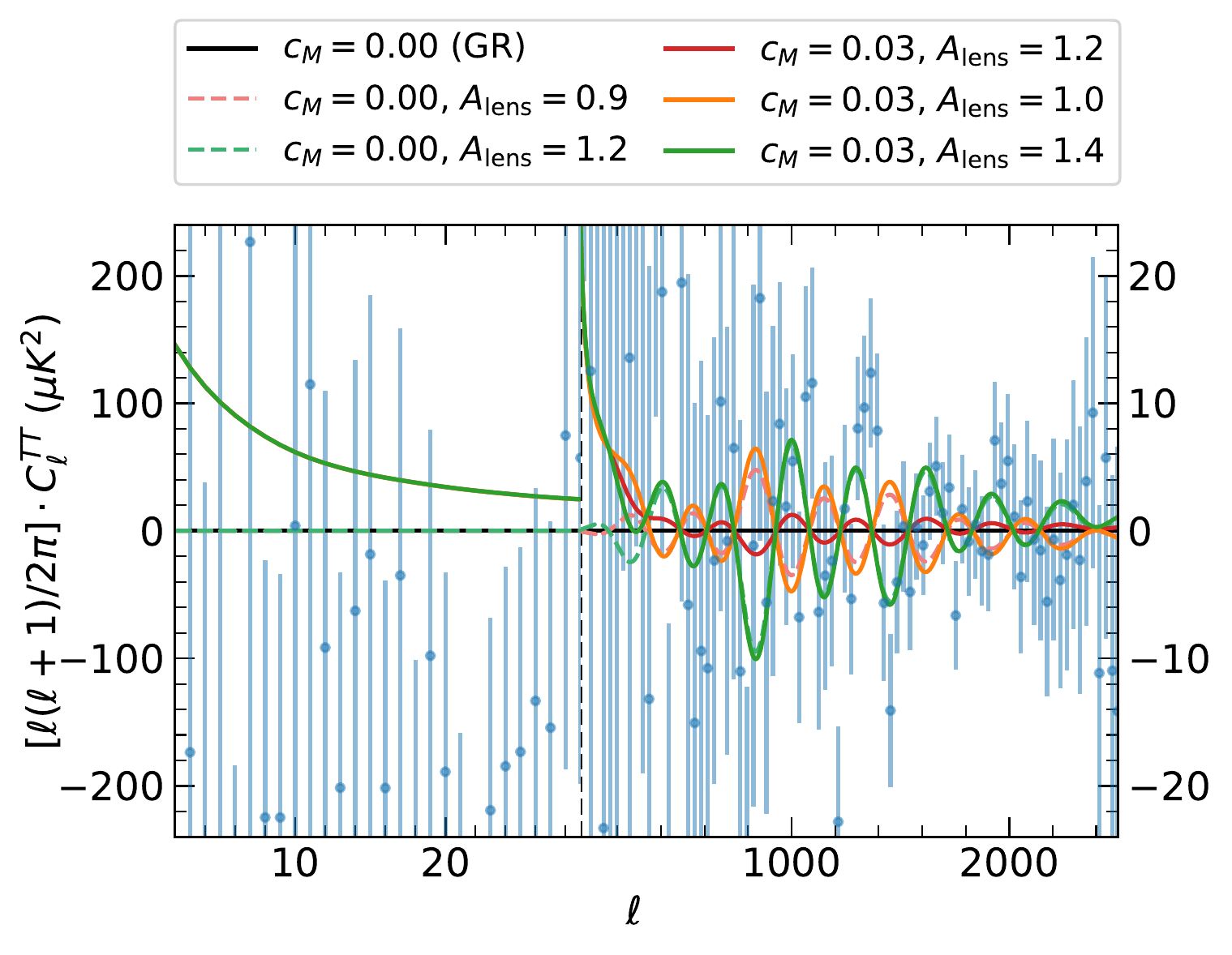} 
\caption{
In comparing the No Slip Gravity model to the Planck 2015 data (points with error bars) 
the lensing suppression from $c_M > 0$ gives a worse fit than GR. However if we allow $\alens >1$ then $c_M > 0$ could be consistent in the acoustic oscillation region. 
We show three pairs of nearly matching curves of such a tradeoff. The fit at low $\ell$ remains discrepant (even with GR, though less so), though the origin of the low power in the data on 
large angular scales is unclear. Note the change from a linear scale in 
$\ell=[2,30]$ to log scale in $\ell>30$, and a change in the vertical scale as well. 
} 
\label{fig:data} 
\end{figure}

Since Planck data prefer a noncanonical (i.e.\ $\Lambda$CDM plus general relativity) $\alens$ with an unknown origin, we can try adding both 
$c_M$ and $\alens$ to the standard six parameters. As we expect from 
Fig.~\ref{fig:phiphimu}, the two can compensate for each other over a broad 
range of multipoles -- but the scale dependence means we cannot satisfy this 
simultaneously at both low and high multipoles. Figure~\ref{fig:data} 
shows the residuals in TT power relative to $\Lambda$CDM plus general relativity 
for three pairs of parameters. Increasing $c_M$ and $\alens$ simultaneously can 
give nearly the same agreement with data in the acoustic peak region. This erases 
the $\ell>30$ penalty in $\chi^2$ from $c_M>0$, but leaves the penalty from 
$\ell<30$. The 95\% confidence upper limit on $c_M$ remains about the same at 0.014.

%%%%%%%%%%%%%%%%%%%%%%%%%%%%%% 

\begin{figure}[htbp!]
\includegraphics[width=\columnwidth]{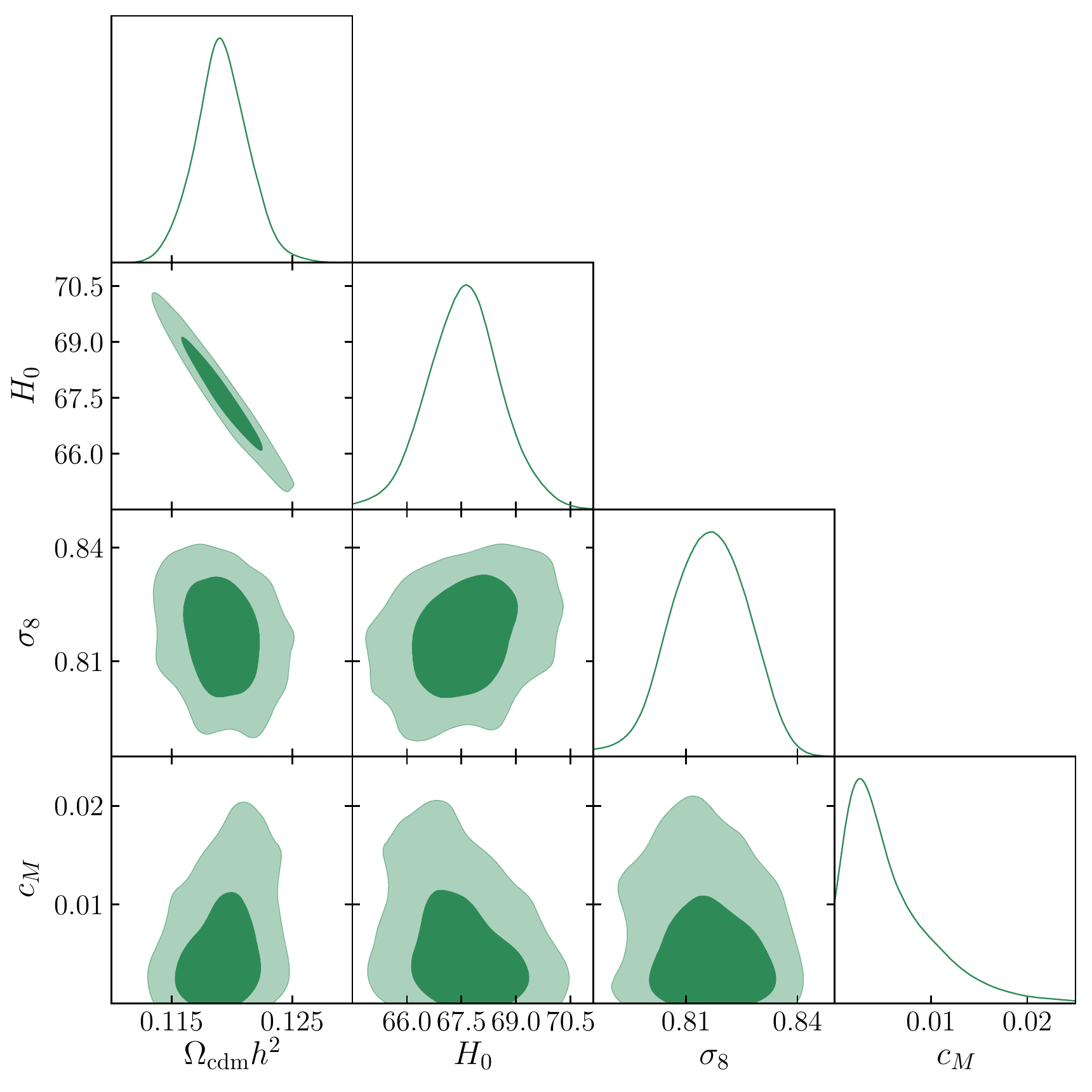} 
\caption{
 The triangle plot for a few select parameters in the MCMC analysis with $\alens=1$. We find that $c_M$ is preferred to be small, and is not covariant with other parameters, even if we allow $\alens$ to vary. 
} 
\label{fig:triangle} 
\end{figure}

We explore the parameter constraints with MCMCs, both fixing $\alens=1$ and allowing it to vary. 
When we fit for it, we obtain $\alens=1.175\pm0.070$, consistent with Planck within general relativity. Since we do not find significant difference in the MCMC results when varying $\alens$, for the figures (except Fig.~\ref{fig:data}) and tables we fix $\alens=1$. The standard cosmological parameters are 
not significantly shifted from their concordance values, as seen in Table~\ref{tb:MCMC}. Figure~\ref{fig:triangle} presents a triangle plot showing four 
of the parameters. We find that $c_M$ has almost no covariance with the other parameters due to its scale dependence.

%%%%%%%%%%%%%%%%%%%%%%%%%%%%%%%%%%%% 
\begin{table}
\begin{tabular}{c|c|c|c}
\hline
Param & Mean$\pm\sigma$ & 95\% lower & 95\% upper \\ \hline
\rule{0pt}{1.1\normalbaselineskip}$10^{2} \Omega_b h^2$ & $2.221_{-0.025}^{+0.025}$ & $2.170$ & $2.274$ \\
\rule{0pt}{1.1\normalbaselineskip}$\Omega_{\rm cdm}h^2$ & $0.1192_{-0.0022}^{+0.0022}$ & $0.1144$ & $0.1238$ \\
\rule{0pt}{1.1\normalbaselineskip}$H_0$ & $67.55_{-1.00}^{+1.03}$ & $65.44$ & $69.84$ \\
\rule{0pt}{1.1\normalbaselineskip}$\sigma_8$ & $0.8159_{-0.0096}^{+0.0117}$ & $0.7955$ & $0.8373$ \\ 
\rule{0pt}{1.1\normalbaselineskip}$c_M$ & $0.0053_{-0.0053}^{+0.0011}$ & $0.0000$ & $0.0153$ 
\end{tabular}
\caption{Results of the MCMC analysis with MontePython and \hic\ for several parameters, keeping $\alens=1$.
}
\label{tb:MCMC}
\end{table}

%%%%%%%%%%%%%%%%%%%%%%%%%%%%%%%%%% 

The Planck data has $A_{\rm lens}=1.180\pm0.065$ \cite{planck6}, as mentioned previously, while the 
Atacama Cosmology Telescope data gives 
$\alens=1.06\pm0.15$(stat)$\pm0.06$(sys) 
\cite{1611.09753}. 
In the case $A_{\rm lens}>1$ from some origin, $c_M > 0$ suppresses the lensing back down and could give a good fit to $\alens=1$. Conversely, the apparent $\alens$ of, e.g., the POLARBEAR CMB experiment \cite{polarbear} with $A_{\rm lens} = 0.60^{+0.26}_{-0.24}\,$(stat)
$^{+0.00}_{-0.04}\,$(inst) $\pm 0.14\,$(foreground) $\pm 0.04\,$(multi) or the South Pole Telescope with $\alens=0.81\pm0.14$ \cite{1707.09353} could be 
interpreted as $\alens=1$ with $c_M>0$. 

Future, high accuracy polarization data, for example from the proposed CMB-S4 experiment, will play an important role in elucidating the hints 
of $\alens$ deviations \cite{1712.08758}. In addition, higher order CMB correlation functions  and  matter power spectrum measurements will place constraints on $\alens$, No Slip Gravity, and general scale dependent physics.

%%%%%%%%%%%%%%%%%%%%%%%%%%%%%%% 
\section{Conclusions} \label{sec:concl} 

No Slip Gravity is similar to the well studied $f(R)$ gravity in that it involves  
only one free function of time in addition to the expansion history. However, while 
$f(R)$ has no effect on light propagation, No Slip Gravity does and in a manner firmly 
connected to its effect on matter growth. That is, there is no slip so $\gm=\gl$. 
Furthermore, due to the strength of its braiding No Slip Gravity can suppress structure 
growth while most scalar-tensor theories enhance it. These properties make it an 
attractive theory to study. 

We focus on the effects on the CMB, in particular the primordial gravitational wave contribution 
to the B mode polarization as well as the lensing B modes. Implementing No Slip Gravity in the Boltzmann code \hic\ (with the modification made publicly available on github at the URL given in Sec.~\ref{sec:method}), we compute the CMB and matter density power spectra. 
Along the way we show that treatment of the kineticity $\al_K$ requires some consideration, 
and we present two ways for successfully treating it. 

The CMB B mode reionization and recombination bumps from primordial gravitational waves 
are affected modestly in No Slip Gravity relative to general relativity, despite the 
equivalence of the gravitational wave speed of propagation, due to the running of the 
Planck mass $\al_M$. We find a fractional change to the tensor to scalar power ratio of 
$\delta r/r\lesssim 0.2\,c_M$, where $c_M$ is maximum value of $\al_M$. For a gravity 
model designed to provide current cosmic acceleration but restore to general relativity 
in the early universe, the main effect is on the reionization bump at low multipoles 
$\ell\lesssim10$. 

The impact on lensing B mode polarization at higher multipoles $\ell\gtrsim100$ is 
much more significant. Here the deviations from general relativity can be of order 
$4c_M$, or tens of percent. We also investigate the CMB lensing power spectrum 
$\clpp$ itself and find a pleasingly accurate simple analytic expression for the 
deviation, based on the integrated effect of $\gm-1$ and the direct relation in No 
Slip Gravity of $\gm$ with $\al_M$ or $M_\star$. We discuss how scale dependence 
enters in $\clpp$ and how it gets washed out in $\clbb$ due to the angular convolution. 
This also distinguishes the CMB lensing amplitude $A_{\rm lens}$ from the B mode amplitude 
$A_{BB}$. 

Finally, we perform a Monte Carlo analysis of our No Slip Gravity model compared to 
the Planck 2015 TT data. 
We find that this CMB data prefers $c_M$ to be smaller than the large scale structure data points to; this comes equally from the higher than $\Lambda$CDM plus general relativity lensing of the CMB ($\alens>1$) and the low large scale power in the CMB. Allowing $A_{\rm lens}>1$ phenomenologically opens up somewhat the parameter space $c_M>0$ though the scale dependence means this is not a 
perfect tradeoff. Future polarization data can shed more light on these issues.

Current data are merely suggestive of a suppression in matter density growth relative 
to the concordance model of general relativity with \lcdm, e.g.\ from redshift space 
distortions or weak lensing, so there is no significant 
evidence for (or against) No Slip Gravity. Next generation CMB experiments such as 
CMB-S4 \cite{cmbs4} and LiteBIRD \cite{litebird} have as a major goal the measurement 
of the B mode polarization power spectrum from the signatures of primordial gravitational 
waves at low multipoles to CMB lensing at higher multipoles. The computations presented 
here enable such observations to be an important 
test of general relativity vs modified gravity. No Slip Gravity is highly predictive, 
and will also be constrained from future galaxy surveys measuring the growth of 
structure, i.e.\ $f\sigma_8(z)$ and weak lensing.
In addition, the theory gives clear relations between standard siren and photon luminosity 
distances to the same redshift that will be tested very precisely by upcoming gravitational wave detectors \cite{Ezquiaga:2018btd}.

%%%%%%%%%%%%%%%%%%%%%% 
\acknowledgments 

This work is supported in part by the Energetic Cosmos Laboratory and by 
the U.S.\ Department of Energy, Office of Science, Office of High Energy 
Physics, under Award DE-SC-0007867 and contract no.\ DE-AC02-05CH11231. 
MZ is supported by the Marie Sklodowska-Curie Global Fellowship Project NLO-CO. 
MB acknowledges the support of the Natural Sciences and Engineering Research Council of Canada (NSERC), [PGSD2-517114-2018].
This research used resources of the National Energy Research Scientific Computing Center, a DOE Office of Science User Facility supported by the Office of Science of the U.S. Department of Energy under Contract No. DE-AC02-05CH11231.

%%%%%%%%%%%%%%%%%%%%%%%%%%%%%%%% 

\end{document}